\documentclass[prl,twocolumn,showpacs,floatfix,amsfonts]{revtex4}
\usepackage{graphicx,graphics,color,epsfig}
\usepackage{bm}
\usepackage{amsmath}
\usepackage{amssymb}
\begin{document}
\preprint{}
\title{Effects of $\tau_1$ Scattering on Fourier-Transformed Inelastic Tunneling Spectra
in High-$T_c$ Cuprates with Bosonic Modes}

\author{Jian-Xin Zhu,$^{1}$ K. McElroy,$^{2,3}$ J. Lee,$^{2}$
T. P. Devereaux,$^{4}$ Qimiao Si,$^{5}$ J. C. Davis,$^{2}$ and  A.
V. Balatsky$^{1}$ }

\affiliation{$^{1}$ Theoretical Division, MS B262, Los Alamos
National Laboratory, Los Alamos, New Mexico 87545, USA}

\affiliation{$^{2}$ LASSP, Department of Physics, Cornell
University, Ithaca, New York 14850, USA}

\affiliation{$^{3}$ Department of Physics, University of California,
Berkeley, California 94720-7300, USA}

\affiliation{$^{4}$ Department of Physics, University of Waterloo,
Ontario, Canada N2L 3GI}

\affiliation{$^{5}$ Department of Physics \& Astronomy, Rice
University, Houston, Texas 77005, USA}

\date{June 21, 2005}
\begin{abstract}
We study the $\tau_1$-impurity induced $\mathbf{q}$-space pattern of
the energy derivative local density of states (LDOS) in a $d$-wave
superconductor. We are motivated in part by the recent scanning
tunneling microscopy (STM) observation of strong gap inhomogeneity
with weak charge density variation in
Bi$_{2}$Sr$_2$CaCu$_2$O$_{8+\delta}$ (BSCCO). The hypothesis is that
the gap inhomogeneity might be triggered by the disorder in pair
potential. We focus on the effects of electron coupling to various
bosonic modes, at the mode energy shifted by the $d$-wave
superconducting gap. The pattern due to a highly anisotropic
coupling of electrons to the $B_{1g}$ phonon mode is similar to
preliminary results from the Fourier transformed inelastic electron
tunneling spectroscopy (FT-IETS) STM experiment in BSCCO. We discuss
the implications of our results in the context of band
renormalization effects seen in the ARPES experiments, and suggest
means to further explore the electron-boson coupling in the
high-$T_c$ cuprates.

\end{abstract}
\maketitle

The extent to which collective excitations of high-$T_c$ cuprates
are manifested in their single particle spectra is a long standing
issue. The band renormalization effects, seen in the
ARPES~\cite{Damascelli03} (as well as in the break junction
tunneling experiments~\cite{Zasadzinski01}), have the
characteristics of an electron-bosonic mode coupling. The ``41 mev''
spin resonance mode, a prominent feature in the spin excitation
spectrum, is a natural candidate for electrons to couple
to~\cite{Norman98,Abanov99,Eschrig00}.
However, there are also phonon modes of similar energies, and they
may instead have the strongest influence on the single electron
spectra~\cite{Lanzara01,Cuk04,Sandvik04,Devereaux04}. At present,
ARPES alone appears inadequate to differentiate the two scenarios.
Sometime ago, several of us proposed~\cite{Zhu04} an FT-IETS STM
technique as a complimentary means to study this issue. The
technique takes advantage of the pioneering work of the Fourier
transformed STM~\cite{Hoffman02b,McElroy03}, and combines it with
the vintage IETS~\cite{Jakievic66,Scalapino67,Balatsky03}.

Central to this technique is the Fourier transform of the energy
derivative of tunneling conductance map in real space
$d^2I/dV^2({\bf r},eV)\rightarrow d^2I/dV^2({\bf q},eV)$. This
$\mathbf{q}$-space map, which can also be called Fourier map,
contains information about inelastic scattering in the system.
Theoretically, one finds peaks in $\bf q$ space and energy $eV$  in
this Fourier map of IETS that are related to the inelastic
scattering off some collective excitations in the system. In the
case of electron-spin mode coupling, the FT-LDOS near an ordinary
potential scattering center (a $\tau_3$ impurity in the Nambu space)
at the energy of $E=\pm (\Delta_0 + \Omega_0)$ was shown to have
sharp features at momenta close to $(\pi,\pi)$. (Here $\Delta_0$ is
the maximum of the $d$-wave superconducting energy gap and
$\Omega_0$ the mode energy.)


Recently preliminary results from the first FT-IETS STM experiment
has been reported in BSCCO~\cite{Lee05}. While  features are
observed in the Fourier-transformed $d^{2}I/dV^{2}$ at the
expected energy range, observed intensity  near $(\pi,\pi)$ is
low. Instead, the strongest intensity appears at the wavevectors
parallel to the Cu-O bond directions. These experimental results
have in turn motivated us to compare the FT-IETS spectra near a
potential scatterer in the cases of electrons coupled to the spin
resonance and various phonon modes~\cite{Zhu05}. It was shown that
all cases contain sharp features near $(\pi,\pi)$, in disagreement
with the experimental spectrum.

This raises the question of whether the $\tau_3$ scatterer correctly
describes the impurities in Bi$_{2}$Sr$_2$CaCu$_2$O$_{8+\delta}$
(BSCCO). The origin of ubiquitously observed nanoscale inhomogeneity
in BSCCO has been investigated by studying the correlation between
the inhomogeneity and the position of oxygen
dopants~\cite{McElroy05}. It is shown that the local electronic
states are not associated with charge density variations. To account
for those features, Nunner {\em et al.}~\cite{Nunner05} proposed to
look at the effect of random pairing potential fluctuations, the
so-called $\tau_1$ disorder. In what follows, we will address the
effect of $\tau_1$ disorder on FT-IETS signatures.

In this Letter, we take the disorder in pair potential as the
scattering center, which is of $\tau_1$ character in the Nambu
space, and study the Fourier component of the energy derivative
local density of states, $d^{2}I/dV^{2}$, at $E \approx
\pm(\Omega_0+\Delta_0)$ around such a scatterer. We considered a few
typical bosonic modes as a possible scattering modes that produce
IETS fingerprints. We find that the results for $\tau_1$ are
qualitatively different from the case of potential disorder: 1)
there are no strong signatures near in  the $\mathbf{q}$-space near
$(\pi,\pi)$ in any of the electron-boson couplings; 2) the highly
anisotropic coupling of electrons to the out-of-plane out-of-phase
oxygen buckling $B_{1g}$ phonon mode, gives rise to a Fourier
pattern similar to the IETS-STM experiment in BSCCO \cite{Lee05}.
Our results are also consistent with the in-plane breathing mode
although the agreement with the data is not as good.

We start with a model Hamiltonian for a two-dimensional $d$-wave
superconductor with the coupling of electrons to bosonic modes:
\begin{equation}
\mathcal{H}=\mathcal{H}_{BCS} + \mathcal{H}_{el-boson} +
\mathcal{H}_{imp} \;.
\end{equation}
Here the bare BCS Hamiltonian,
$\mathcal{H}_{BCS}=\sum_{\mathbf{k},\sigma} \xi_{\mathbf{k}}
c_{\mathbf{k}\sigma}^{\dagger} c_{\mathbf{k}\sigma}
+\sum_{\mathbf{k}}(\Delta_{\mathbf{k}}
c_{\mathbf{k}\uparrow}^{\dagger}c_{-\mathbf{k}\downarrow}^{\dagger}
+\Delta_{\mathbf{k}}^{*}
c_{-\mathbf{k}\downarrow}c_{\mathbf{k}\uparrow})$, where the normal
state energy dispersion is given by~\cite{Norman95},
$\xi_{\mathbf{k}}=-2t_{1} (\cos k_x + \cos k_y) -4t_{2} \cos k_{x}
\cos k_y
 -2t_{3} (\cos 2k_x + \cos 2k_y)
-4t_{4} (\cos 2k_x \cos k_y + \cos k_x \cos 2k_y)
 -4 t_{5} \cos 2k_x \cos 2k_y - \mu$, with $t_1=1$,
$t_{2}=-0.2749$, $t_{3}=0.0872$, $t_4=0.0938$, $t_5=-0.0857$, and
$\mu=-0.8772$, and the $d$-wave gap dispersion
$\Delta_{\mathbf{k}}=\frac{\Delta_{0}}{2}(\cos k_x -\cos k_y)$.
Unless specified explicitly, the energy is measured in units of
$t_1$ hereafter. The coupling of the electrons to bosonic modes is
modeled by the Hamiltonian
$\mathcal{H}_{el-boson}=\frac{1}{\sqrt{N_{L}}}\sum_{\mathbf{k},\mathbf{q}
\atop \sigma} g_{\nu}(\mathbf{k},\mathbf{q})
c_{\mathbf{k}+\mathbf{q},\sigma}^{\dagger}
c_{\mathbf{k}\sigma}(b_{\nu\mathbf{q}} +
b_{\nu,-\mathbf{q}}^{\dagger})$ for the buckling $B_{1g}$ ($\nu=1$)
and the in-plane half breathing ($\nu=2$) modes, while
$\mathcal{H}_{el-boson}=\frac{g_0}{2N_{L}}\sum_{\mathbf{k},\mathbf{q}
\atop \sigma,\sigma^{\prime} }
c_{\mathbf{k}+\mathbf{q},\sigma}^{\dagger}(\mathbf{S}_{\mathbf{q}}\cdot
\bm{\sigma}_{\sigma\sigma^{\prime}}) c_{\mathbf{k},\sigma^{\prime}}$
for the spin resonance mode. For the phonon modes, we consider the
cases where the coupling matrix element is either highly
anisotropic, dependent on both $\mathbf{k}$ and $\mathbf{q}$, or is
only $\mathbf{q}$ dependent. In the following we use the notation
$B_{1g}$-I and $br$-I for the former type of phonon modes while
$B_{1g}$-II and $br$-II for the latter type. Detailed expression of
the coupling matrix elements for these types of phonon modes can be
found in Ref.~\cite{Zhu05}. The third term describes the
quasiparticles scattered off a $\tau_{1}$ impurity due to the
inhomogeneity in pair potential rather than off a conventional
$\tau_{3}$ impurity. In the following, we consider a single
$\tau_{1}$ impurity in a $d-wave$ superconductor. The resulting
Fourier pattern should survive a white-noise random distribution of
such $\tau_1$ impurities in a realistic system. The impurity part of
Hamiltonian can then be written as:
\begin{equation}
\mathcal{H}_{imp}= \delta\Delta \sum_{\delta} \eta_{\delta}
[c_{0\uparrow}^{\dagger}c_{\delta\downarrow}^{\dagger} +
c_{\delta\uparrow}^{\dagger}c_{0\downarrow}^{\dagger} + H.c.]\;,
\end{equation}
where $\eta_{\delta}=1 (-1)$ for $\delta=\hat{x}(\hat{y})$.

To be relevant to recent experimental realization, where no
impurity-induced resonance state was observed, we assume the
$\tau_1$ impurity to have a weak scattering potential
$\delta\Delta$. In this limit, we employ the Born approximation and
arrive at the correction to the LDOS at the $i$-th site, summed over
two spin components:
\begin{equation}
\delta\rho(\mathbf{r}_{i},E)=-\frac{2\delta\Delta}{\pi}\sum_{\delta}
\eta_{\delta}
\mbox{Im}[\hat{\mathcal{G}}(i,0;E+i\gamma)\hat{\tau}_{1}
\hat{\mathcal{G}}(\delta,i;E+i\gamma)]_{11} \;,
\label{ldos_correction}
\end{equation}
where $\hat{\mathcal{G}}$ is the Green's function dressed with the
bosonic renormalization effect and defined in the Nambu
space~\cite{Zhu05}.  From the perspective of the IETS, the energy
derivative of the LDOS, $\frac{d\delta\rho(\mathbf{r}_{i},E)}{dE}$,
is the quantity we are most interested in. It corresponds to the
derivative of the local differential tunneling conductance (i.e.,
$d^{2}I/dV^{2}$). The Fourier component of the differential LDOS is
then given by
$\frac{d\delta\rho(\mathbf{q},E)}{dE}=\sum_{i}\frac{d\delta\rho(\mathbf{r}_{i},E)}{dE}
e^{-i\mathbf{q}\cdot \mathbf{r}_{i}}$ with the spectral weight
defined as $P(\mathbf{q},E)=\biggl{|}
\frac{d\delta\rho(\mathbf{q},E)}{dE}\biggr{|}$.

We consider here for comparison the coupling of electrons to spin
resonance mode, $B_{1g}$ and breathing phonon modes.  For the
numerical calculation, we take the superconducting energy gap
$\Delta_0=0.1$, the frequency of all bosonic modes
$\Omega_{0}=0.15$. The $\tau_1$ impurity scattering strength
$\delta\Delta$ is taken to be 50\% of the superconducting energy
gap. The coupling strength for all types of bosonic modes is
calibrated to give at the Fermi energy $E=0$ an identical frequency
renormalization factor in the self energy. The same procedure as in
Ref.~\cite{Zhu05} is followed to obtain the Fourier spectral weight
$P(\mathbf{q},E)$.

\begin{figure}[th]
\centerline{\psfig{file=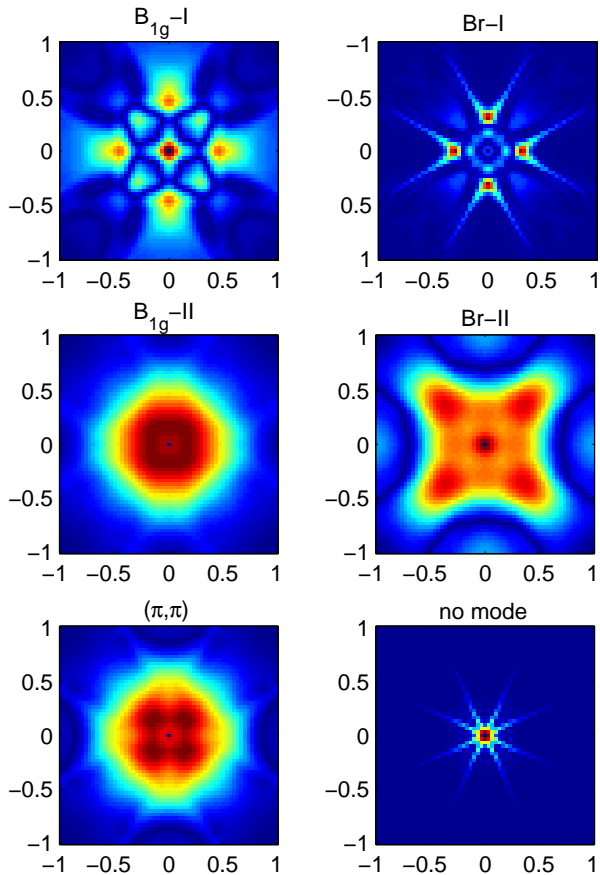,width=8cm}} \caption{The
Fourier spectrum of the energy derivative LDOS at the energy
$E=-(\Delta_0 + \Omega_0)$ for a $d$-wave superconductor with the
electronic coupling to the $B_{1g}$-I, $br$-I, $B_{1g}$-II, $br$-II,
spin resonace modes. For comparison, the spectrum is also shown for
the case of no mode coupling. } \label{FIG:FOURIER-1}
\end{figure}

In Figs.~\ref{FIG:FOURIER-1}, we present the results of the
Fourier spectrum, $P(\mathbf{q},E)$, at the energy $E=-(\Delta_0 +
\Omega_0)$ for a $d$-wave superconductor with the electronic
coupling to the bosonic modes. For comparison, the same spectrum
is also shown (last panel) for the case of no mode coupling. Note
that the case without the mode coupling, the energy $\Omega_0$ has
no special meaning in the context of the electronic properties,
and the energy $E=-(\Delta_0 + \Omega_0)$ is chosen merely for
comparison to the case of mode coupling. The energy $E=-(\Delta_0
+ \Omega_0)$ corresponds to the position where the bosonic modes
are excited, signaling a peak in the IETS $d^2I/dV^2$-$V$
tunneling spectrum~\cite{Zhu05}. First of all, the Fourier maps
for all cases does not display any peak structure at large
  $\mathbf{q}$ near $(\pm\pi,\pm\pi)$ and $(\pm
\pi,\mp\pi)$, which appears persistently with the $\tau_3$
scattering~\cite{Zhu05}. Instead the Fourier spectral weight is
minimal in intensity (dark blue area in the figure) in these
regions. The map for the case of electronic coupling to the
$B_{1g}$-I phonon mode shows locally strongest intensity (red spots)
at $\mathbf{q}$ about $(\pm\frac{2\pi}{4},0)$ and
$(0,\pm\frac{2\pi}{4})$. The peak intensity at $\mathbf{q}$ near
$(\pm \frac{2\pi}{4},\pm \frac{2\pi}{4})$ and $(\pm
\frac{2\pi}{4},\mp\frac{2\pi}{4}$) is much weaker than those along
the bond directions. The map for the case of the coupling to the
$br$-I phonon mode exhibits locally strongest intensity (red spots)
at $\mathbf{q}$ near $(\pm \frac{3\pi}{10},0)$ and $(0,\pm
\frac{3\pi}{10})$. In addition, each of these red spot has a
double-tail structure, which is absent in the case of $B_{1g}$-I
mode coupling. The maps for the cases of the $B_{1g}$-II and spin
resonance mode coupling exhibit similar $\mathbf{q}$ structure. The
finite intensity is uniformly distributed on a circular strip near
$\vert \mathbf{q} \vert=\frac{2\pi}{4}$ and becomes stronger as
$\mathbf{q}$ approaches the zero point. No locally distinguishable
strongest intensity peak can be identified at
$\mathbf{q}=(\pm\frac{2\pi}{4},0)$ and $(0,\pm \frac{2\pi}{4})$. The
map for the coupling to the $br$-II phonon mode exhibits locally the
highest intensity (red spots) at $\mathbf{q}$ near $(\pm
\frac{2\pi}{4},\pm \frac{2\pi}{4})$ and $(\pm \frac{2\pi}{4},
\mp\frac{2\pi}{4})$. No peaks are found at $\mathbf{q}$ near $(\pm
\frac{2\pi}{4},0)$ and $(0,\pm \frac{2\pi}{4})$. The map for the
case of no mode coupling shows an eight-tail star shape at
$\mathbf{q}=(0,0)$, which consists of the head-on overlap of four
red spots, such as those appearing in the case of the $br$-I phonon
mode coupling each with two tails. As we have already emphasized,
experimentally, the Fourier map of $d^2I/dV^2$ shows intensity peaks
only at  $\mathbf{q}=(\pm \frac{2\pi}{5},0)\pm 15\%$ and $(0,\pm
\frac{2\pi}{5}) \pm 15\% $~\cite{Lee05}. Therefore, by comparison
with the experimental data, our new FT-IETS STM analysis also
supports the notion~\cite{Zasadzinski01,Norman98,Abanov99,Eschrig00,
Lanzara01,Cuk04,Sandvik04,Devereaux04} that the electronic band must
be renormalized by its coupling to the bosonic modes. In particular,
the results based on the scenario of highly anisotropic coupling of
electrons to the $B_{1g}$ phonon mode are in best agreement with the
IETS-STM data in BSCCO.

\begin{figure}[th]
\centerline{\psfig{file=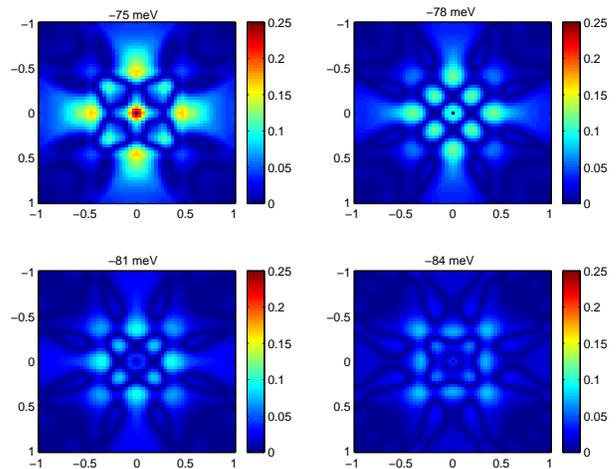,width=8cm}} \caption{The
Fourier spectrum of the derivative of the LDOS is shown at the
various values of the energy for the case of the electronic coupling
to the $B_{1g}$-I phonon mode. Here the energy has been measured by
scaling $\Delta_0=30\;\mbox{meV}$.} \label{FIG:FOURIER-2}
\end{figure}

In Fig.~\ref{FIG:FOURIER-2}, we present the energy evolution Fourier
pattern for the electronic coupling to the $B_{1g}$-I mode. It shows
that the intensity peak structure along the bond direction is robust
against the energy change. The characteristic $q$ vector, at which
the locally highest intensity is located, decreases slightly with
the increased energy. This result is also not inconsistent with the
preliminary experiment.

Our results demonstrate the important role the character of the
scattering center plays in the Fourier spectrum. To explore this
issue further, we note that, in the case of $\tau_1$ scattering
considered here, the Fourier spectrum can be expressed as follows,
\begin{eqnarray}
&\delta\rho(\mathbf{q};E)=\frac{2\delta\Delta}{N_{L}}
\sum_{\mathbf{k}} (\cos k_x - \cos k_y) &
\nonumber \\
&\times  \{ [A(\mathbf{k};E)K(\mathbf{k}+\mathbf{q};E) +
K(\mathbf{k};E)A(\mathbf{k}+\mathbf{q};E)] & \nonumber \\
& +[J(\mathbf{k};E)B(\mathbf{k}+\mathbf{q};E) +
B(\mathbf{k};E)J(\mathbf{k}+\mathbf{q};E)]\}\;,& \label{EQ:FT}
\end{eqnarray}
where
\begin{eqnarray}
A(\mathbf{k};E)&=&-\frac{2}{\pi}
\mbox{Im}[\mathcal{G}_{11}(\mathbf{k};E+i\gamma)]\;,
\label{EQ:A}\\
B(\mathbf{k};E)&=&\mbox{Re}[\mathcal{G}_{11}(\mathbf{k};E+i\gamma)]\;,
\label{EQ:B} \\
J(\mathbf{k};E)&=&-\frac{2}{\pi}
\mbox{Im}[\mathcal{G}_{12}(\mathbf{k};E+i\gamma)]\;,
\label{EQ:J} \\
K(\mathbf{k};E)&=&
\mbox{Re}[\mathcal{G}_{12}(\mathbf{k};E+i\gamma)]\;. \label{EQ:K}
\end{eqnarray}
This expression shows that, for the $\tau_1$ scattering, the Fourier spectrum
is determined by the $\mathbf{k}$-summation of the product terms
constituting the imaginary (real) parts of the single-particle
$(\mathcal{G}_{11})$ with the real (imaginary) parts of anomalous
$(\mathcal{G}_{12})$ Green's function in the superconducting state,
{\em weighted by a $d$-wave type form factor $\cos k_x -\cos k_y$}.
This scattering process with the $\tau_1$ impurity is significantly
different than the case of a $\tau_3$ impurity
scattering~\cite{Zhu05}, where the convolution takes place between
the real and imaginary parts of the same Green's function without
the form factor. This difference of the scattering process matters
significantly in the resulting Fourier map. As shown in
Eq.~(\ref{EQ:FT}), the form factor $\cos k_x -\cos k_y$ appearing in
the $\tau_1$ scattering case is identically zero along the diagonals
in the first Brillouin,
but reaches a maximum at the $M$ points
[$\mathbf{k}=(\pm \pi,0)$ and $(0,\pm \pi)$] on the zone boundary .
It then follows that any stronger intensity from the product of
$AK$, $BJ$ connected by a $\mathbf{q}$ oriented close to the
diagonals is strongly suppressed, while the intensity connected by
$\mathbf{q}$ oriented parallel to the bond direction is enhanced.
For the electronic coupling to the $B_{1g}$-I phonon mode, it has
been found~\cite{Zhu05} that there are moderately strong intensity
on the two split beams oriented perpendicular to the zone boundary
at M points in the function $A$, $B$, $J$, and $K$. The form factor
$\cos k_x -\cos k_y$ then tips the relative contribution from the
product $AK$ and $BJ$, giving rise to locally highest intensity at
$\mathbf{q}=(\pm\frac{2\pi}{4},0)$ and $(0,\pm \frac{2\pi}{4})$ in
the Fourier map (see the first panel of Fig.~\ref{FIG:FOURIER-1}).
These split beams of intensity are absent for the electronic
coupling to other modes.

Our results naturally suggest additional means to further explore the
electron-bosonic mode coupling experimentally. For instance, a Zn impurity
acts as a nonmangetic
potential center -- a $\tau_3$ scatterer. In this case, the sharp
features near $(\pi,\pi)$ should be observed in the FT-IETS spectrum.
In the case of low-energy elastic scattering interference of
quasiparticles, related effects have in fact been demonstrated.
Indeed, strong signatures near $(\pm\pi,\pm\pi)$ and $(\pm \pi,\mp\pi)$
appear in the theoretical spectra near a $\tau_3$ scatterer~\cite{Wang03}.
These features are not observed experimentally in the stoichiometrical
BSCCO~\cite{Hoffman02b,McElroy03}, but are seen around a nomagnetic
Zn impurity in the doped BSCCO.

In conclusion, we have studied, for the first time, the
$\tau_1$-impurity induced Fourier pattern of the energy derivative
local density of states in a $d$-wave superconductor with the
coupling of electrons to various bosonic modes. We consider
$B_{1g}$, half-breathing, and spin $(\pi,\pi)$ modes. Our results
show that, at the mode energy shifted by the $d$-wave
superconducting gap energy $\Delta_0$, i.e., $E=\pm(\Delta_0 +
\Omega_0)$, the coupling of electrons to the $B_{1g}$ or breathing
phonon modes, gives rise to a Fourier pattern similar to the
preliminary Fourier transformed IETS-STM experiment in
BSCCO~\cite{Lee05}. The coupling of electrons to the spin resonance
mode, on the other hand, yields a Fourier spectrum that is
inconsistent with the experiment. These results  do not necessarily
rule out the role of the spin-spin interactions as being relevant
for superconductivity in BSCCO, instead they imply that
electron-phonon coupling has a strong impact on the superconducting
electronic structure. These results have important implications for
our understanding of the electronic properties of the cuprates. They
also demonstrate the potential of the FT-IETS STM technique, and
highlight the importance of $\tau_1$ scattering in the impurity-free
BSCCO~\cite{Nunner05}.

We thank D.-H. Lee, N. Nagaosa, M. R. Norman, D. J. Scalapino, and
Z. X. Shen for very useful discussions. This work was supported by
the US DOE (J.X.Z. and A.V.B.), the NSERC, the Office of Naval
Research under Grant No. N00014-05-1-0127, and the A. von Humboldt
Foundation (T.P.D.), the NSF under Grant No. DMR-0424125 and the
Robert A. Welch Foundation (Q.S.), the Office of Naval Research
under grant N00014-03-1-0674, the NSF under Grant No. DMR-9971502,
the NSF-ITR FDP-0205641, and the Army Research Office under Grant
No. DAAD19-02-1-0043 (K.M., J.L., and J.C.D).


\begin{thebibliography}{99}

\bibitem{Damascelli03} A. Damascelli, Z. Hussain, and Z.-X. Shen, Rev.
Mod. Phys. {\bf 75}, 473 (2003), and references therein.

\bibitem{Zasadzinski01} J. F. Zasadzinski, L. Ozyuzer, N. Miyakawa,
K. E. Gray, D. G. Hinks, and C. Kendziora, Phys. Rev. Lett. {\bf
87}, 067005 (2001), and references therein.

\bibitem{Norman98}
M. R. Norman, M. Eschrig, A. Kaminski, and J. C. Campuzano, Phys.
Rev. B {\bf 64}, 184508 (2001); M. R. Norman and H. Ding, Phys. Rev.
B {\bf 57}, 11089 (1998).

\bibitem{Abanov99} Ar. Abanov and A. V. Chubukov, Phys. Rev. Lett.
{\bf 83}, 1652 (1999).

\bibitem{Eschrig00} M. Eschrig and M. R. Norman, Phys. Rev. Lett.
{\bf 85}, 3261 (2000).

\bibitem{Lanzara01} A. Lanzara, P. V. Bogdanov, X. J. Zhou, S. A. Kellar,
D. L. Feng, E. D. Lu, T. Yoshida, H. Eisaki, A. Fujimori, K. Kishio,
J.-I. Shimoyama, T. Noda, S. Uchida, Z. Hussain, Z.-X. Shen, Nature
{\bf 412}, 510 (2001).

\bibitem{Cuk04}  T. Cuk, F. Baumberger, D. H. Lu, N. Ingle,
X. J. Zhou, H. Eisaki, N. Kaneko, Z. Hussain, T. P. Devereaux, N.
Nagaosa, and Z.-X. Shen, Phys. Rev. Lett. {\bf 93}, 117003 (2004).

\bibitem{Sandvik04} A. W. Sandvik, D. J. Scalapino, and N. E.
Bickers, Phys. Rev. B {\bf 69}, 094523 (2004).

\bibitem{Devereaux04} T. P. Devereaux, T. Cuk, Z.-X. Shen, and N.
Nagaosa, Phys. Rev. Lett. {\bf 93}, 117004 (2004).

\bibitem{Zhu04} J.-X. Zhu, J. Sun, Q. Si, and A. V. Balatsky,
Phys. Rev. Lett. {\bf 92}, 017002 (2004).

\bibitem{Hoffman02b} J. E. Hoffman, K. McElroy,
D.-H. Lee, K. M. Lang, H. Eisaki, S. Uchida, and J. C. Davis,
Science {\bf 297}, 1148 (2002).

\bibitem{McElroy03} K. McElroy, R. W. Simmonds,
J. E. Hoffman, D. H. Lee, J. Orenstein, H. Eisaki, S. Uchida, J. C.
Davis, Nature {\bf 422}, 592 (2003).


\bibitem{Jakievic66} R.C. Jaklevic and J. Lambe, Phys. Rev. Lett. {\bf
17}, 1139 (1966).

\bibitem{Scalapino67} D.J. Scalapino and S.M. Markus, Phys. Rev. Lett. {\bf 18}, 459
(1967).

\bibitem{Balatsky03} A.V. Balatsky,  Ar. Abanov, and J.-X. Zhu,
Phys. Rev. B {\bf 68}, 214506 (2003).

\bibitem{Lee05}  J. Lee, K. McElroy, J. Slezak, S. Uchida, H. Eisaki, and J.C. Davis,
Bull. Am. Phys. Soc. {\bf 50}, 299, (2005); J.C. Davis, J. Lee, K.
McElroy, J. Slezak, H. Eisaki, and S. Uchida, Bull. Am. Phys. Soc.
{\bf 50}, p1223, (2005), J. Lee {\it et al}, unpublished.

\bibitem{Zhu05} J.-X. Zhu {\em et al.}, cond-mat/0507610
(unpublished).

\bibitem{McElroy05} K. McElroy {\em et al.}, to appear in Science.

\bibitem{Nunner05} T. S. Nunner, B. M. Anderson, A. Melikyan, and P.
J. Hirschfeld, cond-mat/0504693 (unpublished).

\bibitem{Norman95} M. R. Norman, M. Randeria, H. Ding, and J. C.
Campuzano, Phys. Rev. B {\bf 52}, 615 (1995).

\bibitem{Wang03}  Q.-H. Wang and
D.-H. Lee, Phys. Rev. B {\bf 67}, 020511 (2003).
\end{thebibliography}
\end{document}